\documentclass{vgtc}                          % final (conference style)
\ifpdf%                                % if we use pdflatex
  \pdfoutput=1\relax                   % create PDFs from pdfLaTeX
  \usepackage{graphicx}                % allow us to embed graphics files
  \DeclareGraphicsExtensions{.pdf,.png,.jpg,.jpeg} % for pdflatex we expect .pdf, .png, or .jpg files
\else%                                 % else we use pure latex
  \usepackage{graphicx}                % allow us to embed graphics files
  \DeclareGraphicsExtensions{.eps}     % for pure latex we expect eps files
\fi%

\graphicspath{{figures/}{pictures/}{images/}{./}} % where to search for the images

\usepackage{microtype}                 % use micro-typography (slightly more compact, better to read)
\PassOptionsToPackage{warn}{textcomp}  % to address font issues with \textrightarrow
\usepackage{textcomp}                  % use better special symbols
\usepackage{mathptmx}                  % use matching math font
\usepackage{times}                     % we use Times as the main font
\usepackage{cite}                      % needed to automatically sort the references
\usepackage{tabu}                      % only used for the table example
\usepackage{booktabs}                  % only used for the table example
\usepackage{setspace}
\usepackage{amsmath}

\newcommand{\etal}{\textit{et al.}}
\newcommand{\ie}{\textit{i.e.}}
\newcommand{\eg}{\textit{e.g.}}
\newcommand{\etc}{\textit{etc.}}
\DeclareMathAlphabet{\mathcal}{OMS}{cmsy}{m}{n}

\onlineid{1030}

\vgtccategory{Research}

\title{Symmetrical Reality: Toward a Unified Framework \\for Physical and Virtual Reality}

\author{Zhenliang Zhang\thanks{e-mail: zzlyw10@gmail.com}\\ %
        \scriptsize Beijing Institute of Technology %
\and Dongdong Weng\thanks{e-mail: crgj@bit.edu.cn}\\ %
     \scriptsize Beijing Institute of Technology %
\and Yue Liu\thanks{e-mail: liuyue@bit.edu.cn}\\ %
     \scriptsize Beijing Institute of Technology
\and Yongtian Wang\thanks{e-mail: wyt@bit.edu.cn}\\ %
     \scriptsize Beijing Institute of Technology
     }
     
\author{
Zhenliang Zhang$^{1}$, Cong Wang$^{3}$, Dongdong Weng$^{1,2,}$\thanks{crgj@bit.edu.cn}, Yue Liu$^{1,2}$, Yongtian Wang$^{1,2}$ \\
    \scriptsize 
    \leftline{
    \lbrack 1\rbrack\ Beijing Engineering Research Center of Mixed Reality and Advanced Display, Beijing Institute of Technology, Beijing, China}\\ 
    \scriptsize 
    \leftline{
    \lbrack 2\rbrack\ AICFVE of Beijing Film Academy, 4 Xitucheng Rd, Haidian, Beijing, China}\\
    \scriptsize 
    \leftline{
    \lbrack 3\rbrack\ China Electronics Standardization Institute, Beijing, China}
    }   

\abstract{
In this paper, we review the background of physical reality, virtual reality, and some traditional mixed forms of them. Based on the current knowledge, we propose a new unified concept called symmetrical reality to describe the physical and virtual world in a unified perspective. Under the framework of symmetrical reality, the traditional virtual reality, augmented reality, inverse virtual reality, and inverse augmented reality can be interpreted using a unified presentation. We analyze the characteristics of symmetrical reality from two different observation locations (\ie, from the physical world and from the virtual world), where all other forms of physical and virtual reality can be treated as special cases of symmetrical reality. 

} % end of abstract

\CCScatlist{
  \CCScatTwelve{Human-centered computing}{Human computer interaction (HCI)}{Interaction paradigms}{Mixed/augmented reality}
}

\begin{document}


\firstsection{Introduction}

\maketitle

It has been a long time since the first concept of virtual reality (VR) was proposed. Actually, that is the very beginning for our physical world to embrace the virtual world (or the digital world). As of now, we have heard a lot of concepts about various kinds of forms that represent the combination of the physical world and the virtual world, such as virtual reality, mixed reality (MR) \cite{milgram1994taxonomy}, augmented reality (AR) \cite{caudell1992augmented}, inverse virtual reality (IVR)\cite{zhang2018Inverse}, and inverse augmented reality (IAR)\cite{zhang2018inverseaug}. Each form has its specific application field, but all of these forms share a common framework. 

\textit{Symmetrical Reality (SR)} is to unify all of the existing forms about the combination of the physical world and the virtual world. Symmetrical reality is defined regarding two crucial attributes: \textit{Object} and \textit{Interaction}. For \textit{Object}, the ratio between the virtual elements and the physical elements determines the existing style of symmetrical reality. For example, if all elements are virtual, SR appears to be virtual reality, and if both virtual parts and physical parts exist, SR appears to be the mixed reality. For \textit{Interaction}, two kinds of actors should be considered. If the actor is located in the physical world, SR appears to be the traditional virtual reality or augmented reality. Conversely, if the actor is located in the virtual world, SR appears to be the inverse virtual reality or inverse augmented reality.

We all know that there is still some undiscovered matter of our world. Inspired by the physical concept of the parallel world \cite{berezhiani2004mirror}, the virtual world might be kind of similar to a parallel world of the physical world. If we consider the physical world as a combination of the humans and the environment, the virtual world should also be composed of the similar two components, \ie, the virtual agents and the virtual environment. Since the physical environment can evolve by itself and humans are fully autonomous individuals, it is expected that the virtual environment and the virtual agent should also possess the capability of evolving automatically. Artificial intelligence can be used to make the virtual environment more intelligent \cite{luck2000applying}, and the virtual world, including virtual agents, can even evolve by itself. Hence the virtual environment and the virtual agent can be treated as two kinds of agents \cite{mateus2016intelligent}, \ie, the character agent and the virtual environment agent. Here we notice that the virtual world is created by humans, but it can grow according to its own rules without direct intervention from the physical world, though the physical world can still send some signals to the virtual world.

\begin{figure}[t!]
    \centering
    \includegraphics[width=\linewidth]{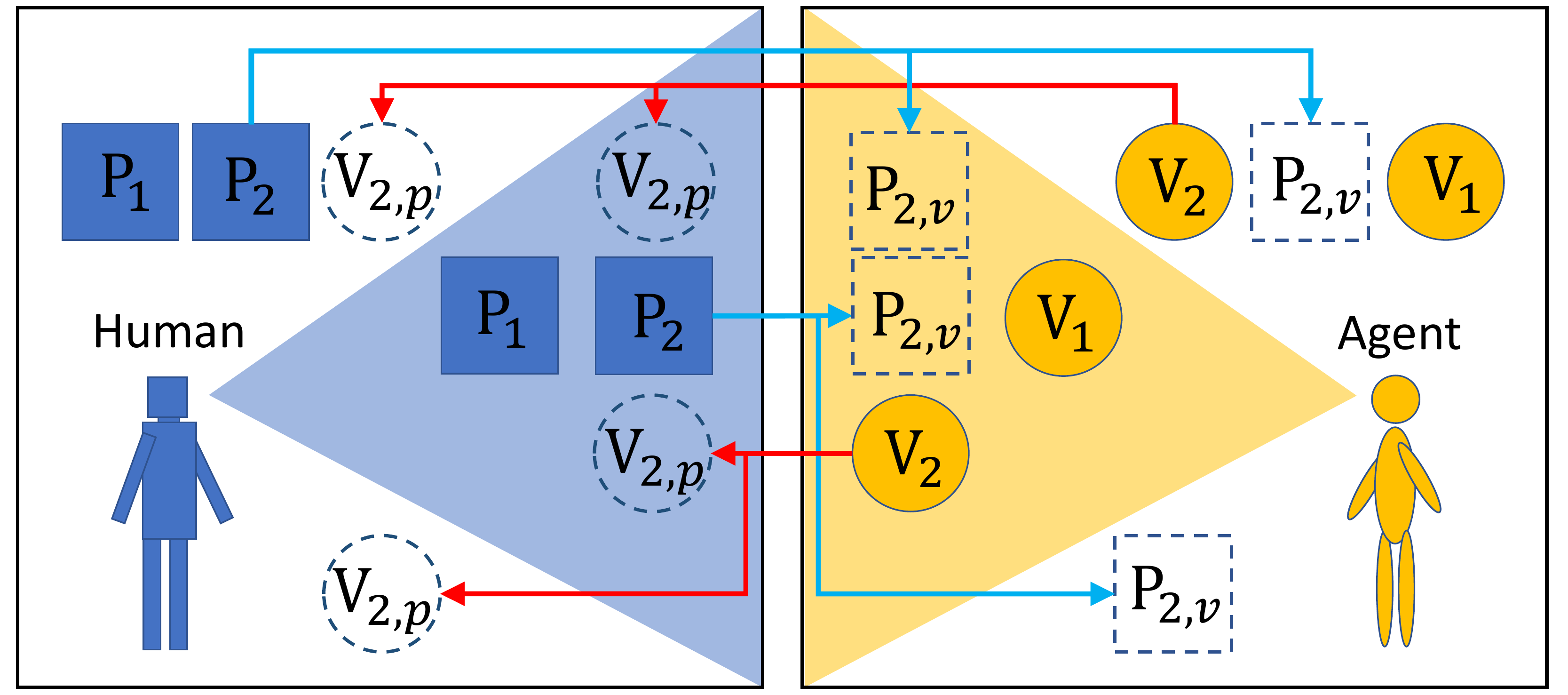}
    \vspace{-21pt}
    \caption{
    \setstretch{0.9}
    Basic structure of symmetrical reality. (Left) The physical world is represented by blue elements. The physical human can sense a part of  $P_1, P_2$ and $V_{2,p}$. (Right) The virtual world is represented by orange elements. The virtual agent can sense a part of $V_1, V_2$ and $P_{2,v}$. }\vspace{-15pt} 
    \label{fig:firstpage}
\end{figure}

Some researchers have proposed related works belonging to the proposed SR framework. For example, the concept of ``dual reality''~\cite{lifton2009dual} refers to building the reflected virtual and physical world, which is similar to the mutually mirrored world \cite{zhang2018Inverse} regarding the structure. However, the former one does not consider a self-learning ability of the virtual agent when compared to the later one. In addition, Roo \etal \cite{roo2017one} proposed the ``one reality'' system, which contained a 6-level mixture of virtual and real contents ranging from purely physical to purely virtual world, which is also a human-centered framework. 
The above contents are discussing the \textit{Object} attribute of SR, and the other attribute called \textit{Interaction} is more significant under this framework. We emphasize the equivalent interaction between the physical world and the virtual world. Milgram \etal \cite{milgram1994taxonomy} analyzed the concepts of AR, MR, and VR, which treated the human as the absolute center of the whole framework. However, the proposed SR can extend the prior framework by adding the virtual agent-centered part \cite{zhang2018inverseaug}. Therefore, the proposed SR break the current frontier by considering a two-center symmetrical world structure and the symmetrical interaction style between the physical and virtual world.
In this paper, we proposed the framework of symmetrical reality to unify the current concepts related to physical and virtual reality.

\section{Theory of Symmetrical Reality}
The typical structure of SR is shown in Fig.~\ref{fig:firstpage}. All solid squares ($P_1, P_2$) denote physical objects in the physical world, and all empty squares ($P_{2,v}$) denote the virtual correspondences of the physical objects. Symmetrically, all solid circles ($V_1, V_2$) denote virtual objects in the virtual world, and all empty circles ($V_{2,p}$) denote the physical correspondences of the virtual objects.

Let $W_p$ denote the set of objects that exist in the physical world, including the virtual-to-physical mapping objects of some virtual objects, $W_v$ denote the set of objects that exist in the virtual world, including the physical-to-virtual mapping objects of some physical objects, then we have

\vspace{-6pt}
\begin{equation}  
\left\{  
    \begin{array}{lr}
        W_p = P_1 \cup P_2 \cup V_{2,p} \\  
        W_v = V_1 \cup V_2 \cup P_{2,v}
    \end{array}
\right.,   \vspace{-6pt}
\end{equation}
where $P_1$ denote the set of physical objects that are not sensed by the physical human, $P_2$ the set of physical objects that has been sensed by the physical human, $V_{2,p}$ the set of the virtual-to-physical mapping objects that are originally located in the virtual world. Similarly, we can know the definitions of $V_1$, $V_2$ and $ P_{2,v}$ by replacing ``physical'' with ``virtual''.
Let $W_{SR}$ denote the set of objects in symmetrical reality, then 

\vspace{-6pt}
\begin{equation}
    W_{SR} = W_p \cup W_v \cup H_p \cup H_v, \vspace{-6pt}
\end{equation}
where $H_p$ and $H_v$ denote the set of physical humans and the set of virtual humans (or virtual agent), respectively. The physical humans are not treated as general physical objects, because they have independent minds and they can be observers of the SR system. The same thing happens to the virtual humans.

We define an operation ``$\star$'' on the set $W_{SR}$. For two given sets $a$ and $b$, which are subsets of $W_{SR}$, $a \star b$ denote the status that $a$ and $b$ coexist in one system. Since in an SR system, the physical human and the virtual human both have the ability to observe the outer environment, we can define the observation function for both of them. Let $F_{H_p}(\cdot)$ denote the observation function of the physical human, and $F_{H_v}(\cdot)$ denote the observation function of the virtual human. The input of the functions can be any subset of $W_{SR}$, and the output of the functions are the observed part of the input set. For example, if there are input sets $I, I_1, I_2$, we have the following rules: $F_{H_p}(I) = [I;\alpha]$ and $F_{H_p}(I_1 \cup I_2) = [I_1;\alpha_1]\star [I_2;\alpha_2]$ , which means that the percentage of $I, I_1, I_2$ being observed is $\alpha, \alpha_1, \alpha_2$, respectively. There should be
\vspace{-6pt}
\begin{equation}  
    \begin{aligned}
    F_{H_p}(W_p)&= F_{H_p}(P_1 \cup P_2 \cup V_{2,p})\\
    &= [P_1;\alpha_1]\star[P_2;\alpha_2]\star[V_{2,p};\alpha_3],
    \end{aligned}\vspace{-6pt}
\end{equation}
and
\begin{equation}  
    \begin{aligned}
    F_{H_v}(W_v)&= F_{H_v}(V_1 \cup V_2 \cup P_{2,v})\\
    &= [V_1;\beta_1]\star[V_2;\beta_2]\star[P_{2,v};\beta_3],
    \end{aligned} \vspace{-3pt}
\end{equation}
where $\alpha_1, \alpha_2, and \alpha_3$ indicate the ratio by which the $P_1, P_2, and V_{2,p}$ are observed by the physical human, and $\beta_1, \beta_2, and \beta_3$ indicate the ratio by which the $V_1, V_2, and P_{2,v}$ are observed by the virtual human.
The model of the proposed SR framework can be represented as:
\vspace{-6pt}
\begin{equation}  \label{eq:sense}
    \begin{aligned}
    \mathcal{M}_{SR} =& F_{H_p}(W_{SR})\star F_{H_v}(W_{SR})\\
    =& F_{H_p}(W_p \cup W_v \cup H_p \cup H_v)\star F_{H_v}(W_p \cup W_v \cup H_p \cup H_v)\\
    =& F_{H_p}(W_p \cup H_v)\star F_{H_v}( W_v \cup H_p )\\
    =& F_{H_p}(P_1 \cup P_2 \cup V_{2,p} \cup H_v)\star F_{H_v}( V_1 \cup V_2 \cup P_{2,v}\cup H_p)\\
    =& [P_1;\alpha_1]\star[P_2;\alpha_2]\star[V_{2,p};\alpha_3] \star [H_v; \alpha_4] \star [V_1;\beta_1]\star[V_2;\beta_2]\\
    &\star[P_{2,v};\beta_3] \star [H_p;\beta_4], 
    \end{aligned}
\end{equation}
where we hold an assumption that the physical human cannot sense the $W_v$ and the virtual human cannot sense the $W_p$. In Eq.~\ref{eq:sense}, to simplify the discussion, we also assume that the physical human does not sense himself/herself, and the virtual human does not sense itself, either.

\begin{table}[b]
\vspace{-12pt}
\caption{
\setstretch{0.9}
Parameter configuration of typical modes of the combination of the virtual world and the physical world. The bottom row indicates that all parameters can range from 0 to 1 in SR, while some parameters cannot be 0 in other modes.}
\vspace{-3pt}
    \centering
    \small
    \renewcommand{\arraystretch}{0.8}
    \begin{tabu}{X[c]cccccccc}
    \toprule
    Mode  & $\alpha_1$ & $\alpha_2$ & $\alpha_3$ & $\alpha_4$ & $\beta_1$ & $\beta_2$ & $\beta_3$ & $\beta_4$ \\
    \midrule
    VR    & 0 & 0 & (0,1] & (0,1] & 0 & 0 & 0 & 0 \\
    AR    & (0,1] &(0,1]  & (0,1] &(0,1]  & 0 & 0 &0  & 0 \\
    IVR   & 0 &0  & 0 & 0 &0  & 0 &(0,1]  & (0,1] \\
    IAR   & 0 & 0 & 0 &0  & (0,1] & (0,1] & (0,1] & (0,1] \\
    SR    & [0,1] &[0,1]  & [0,1] &[0,1]  &[0,1]  & [0,1] &[0,1]  & [0,1] \\
    \bottomrule
    \end{tabu}
    \label{tab:my_table}
\end{table}

\section{Special Cases of Symmetrical Reality}

According to Eq.~\ref{eq:sense}, if we set the parameters to different values, the SR will converge to different special cases, as shown in Table~\ref{tab:my_table}. In this table, all parameters can range from 0 to 1. For example, if $\alpha_1$ equals 0, it means $P_1$ cannot be sensed by the physical human at all; if $\alpha_1$ is a greater than 0, it means $P_1$ can be sensed by the physical human. Note that the parameters only determine the ratio that the objects are sensed by physical or virtual humans. No matter what values the parameters are, all physical and virtual objects, including the mapping relations between the physical and the virtual world, always exist in the symmetrical reality.

\vspace{-3pt}
\paragraph{\textbf{Virtual Reality: Physical Human's Perspective}}
Virtual reality is a paradigm which treats the physical human as the center of the whole system. In a VR system, the human need to wear a kind of device (\eg, the immersive head-mounted display) to experience the virtual vision and even other virtual feelings.
\vspace{-3pt}
\paragraph{\textbf{Augmented Reality: Physical Human's Perspective}}
Similar to VR, augmented reality \cite{caudell1992augmented} is a paradigm which also treats the physical human as the center of the whole system. Different from a VR system, the AR system can provide the human with a scene that mixed both the virtual and the real elements. Another concept called mixed reality is similar to the above structure, but it emphasizes the high fusion of the virtual and the real elements.
\vspace{-3pt}
\paragraph{\textbf{Inverse Virtual Reality: Virtual Agent's Perspective}}
Inverse virtual reality \cite{zhang2018Inverse} is a paradigm which treats the virtual agent (or virtual human) as the center of the whole system. In an IVR system, the virtual agent, which may be created by humans but can develop by itself, acts as a physical human regarding the living habits. Hence it could also sense all things in the physical world (\eg, the physical objects, and the physical human, \etc) that surrounds it.
\vspace{-3pt}
\paragraph{\textbf{Inverse Augmented Reality: Virtual Agent's Perspective}}
Similar to IVR, inverse augmented reality \cite{zhang2018inverseaug} is a paradigm which also treats the virtual agent (or virtual human) as the center of the whole system. Different from an IVR system, the IAR system can provide the virtual agent a mixed scene of the world. Specifically, all the physical elements become virtual, while all the virtual elements become kind of ``real'' because the virtual agent itself is exactly a virtual object.

\section{Conclusion and Future Work}
Symmetrical reality is a unified concept of the combination of the virtual world and the real world. We derive the formulation of SR in the perspective of set theory, which may provide some insight for understanding and designing various systems that bridge the virtual world and the real world. Especially, the artificial intelligence is assumed to construct the mind of the virtual agent in the virtual world. The understanding of SR should be helpful in the future exploration about mixed space of the virtuality and the reality, where the human intelligence and the artificial intelligence coexist.

\acknowledgments{
This work was supported by the National Key Research and Development Program of China (No. 2017YFB1002504) and the National Natural Science Foundation of China (No. U1605254).}

\setstretch{0.917}
\bibliographystyle{abbrv-doi}

\bibliography{template}

\begin{thebibliography}{1}

\bibitem{berezhiani2004mirror}
Z.~Berezhiani.
\newblock Mirror world and its cosmological consequences.
\newblock {\em International Journal of Modern Physics A}, 19(23):3775--3806,
  2004.

\bibitem{caudell1992augmented}
T.~P. Caudell and D.~W. Mizell.
\newblock Augmented reality: An application of heads-up display technology to
  manual manufacturing processes.
\newblock In {\em System Sciences, 1992. Proceedings of the Twenty-Fifth Hawaii
  International Conference on}, vol.~2, pp. 659--669. IEEE, 1992.

\bibitem{lifton2009dual}
J.~Lifton and J.~A. Paradiso.
\newblock Dual reality: Merging the real and virtual.
\newblock In {\em International Conference on Facets of Virtual Environments},
  pp. 12--28. Springer, 2009.

\bibitem{luck2000applying}
M.~Luck and R.~Aylett.
\newblock Applying artificial intelligence to virtual reality: Intelligent
  virtual environments.
\newblock {\em Applied Artificial Intelligence}, 14(1):3--32, 2000.

\bibitem{mateus2016intelligent}
S.~Mateus and J.~Branch.
\newblock Intelligent virtual environment using a methodology oriented to
  agents.
\newblock In {\em International Conference on Virtual, Augmented and Mixed
  Reality}, pp. 714--723. Springer, 2016.

\bibitem{milgram1994taxonomy}
P.~Milgram and F.~Kishino.
\newblock A taxonomy of mixed reality visual displays.
\newblock {\em IEICE Transactions on Information and Systems},
  77(12):1321--1329, 1994.

\bibitem{roo2017one}
J.~S. Roo and M.~Hachet.
\newblock One reality: Augmenting how the physical world is experienced by
  combining multiple mixed reality modalities.
\newblock In {\em Proceedings of the 30th Annual ACM Symposium on User
  Interface Software and Technology}, pp. 787--795. ACM, 2017.

\bibitem{zhang2018Inverse}
Z.~Zhang, B.~Cao, J.~Guo, D.~Weng, Y.~Liu, and Y.~Wang.
\newblock Inverse virtual reality: intelligence-driven mutually mirrored world.
\newblock In {\em Proceedings of IEEE Conference on Virtual Reality and 3D User
  Interfaces (VR)}, pp. 735--736. IEEE, 2018.

\bibitem{zhang2018inverseaug}
Z.~Zhang, D.~Weng, H.~Jiang, Y.~Liu, and Y.~Wang.
\newblock Inverse augmented reality: A virtual agent's perspective.
\newblock In {\em Proceedings of International Symposium on Mixed and Augmented
  Reality (ISMAR)}, 2018.

\end{thebibliography}
\end{document}